\documentclass[fleqn,12pt,twoside]{article}
\usepackage{espcrc1}
\usepackage{graphicx}

\usepackage{floatflt}

\newcommand{\snn}{\sqrt{s_{NN}}}
\newcommand{\seff}{\s_{\rm eff}}
\newcommand{\s}{\sqrt{s}}
\newcommand{\pp}{pp}
\newcommand{\pbarp}{\overline{p}p}

\newcommand{\epem}{e^+e^-}

\newcommand{\nch}{N_{ch}}
\newcommand{\np}{N_{part}}

\newcommand{\ntot}{\langle\nch\rangle}
\newcommand{\avenp}{\langle\np\rangle}

\newcommand{\halfnp}{\langle\np/2\rangle}
\newcommand{\etap}{\eta^{\prime}}

\newcommand{\etaone}{|\eta| < 1}
\newcommand{\dndeta}{d\nch/d\eta}
\newcommand{\dndetap}{d\nch/d\etap}

\newcommand{\dndetaone}{\dndeta|_{\etaone}}

\newcommand{\dndetapnp}{\dndetap / \halfnp}

\newcommand{\ratio}{\ntot/\halfnp}

\newcommand{\nubar}{\overline{\nu}}
\newcommand{\yb}{y_{\rm beam}}

\title{Universal Behavior of Charged Particle Production 
in Heavy Ion Collisions}

\author{Peter A. Steinberg$^{2}$ 
\thanks{Current address: Physics Department, University of Cape Town, South Africa}
for the PHOBOS collaboration\\
\vspace*{1mm}
{
\footnotesize
B.B.Back$^1$,
M.D.Baker$^2$,
D.S.Barton$^2$,
R.R.Betts$^6$,
M.Ballintijn$^4$,
A.A.Bickley$^7$,
R.Bindel$^7$,
A.Budzanowski$^3$,
W.Busza$^4$,
A.Carroll$^2$,
M.P.Decowski$^4$,
E.Garc\'{i}a$^6$,
N.George$^{1,2}$,
K.Gulbrandsen$^4$,
S.Gushue$^2$,
C.Halliwell$^6$,
J.Hamblen$^8$,
G.A.Heintzelman$^2$,
C.Henderson$^4$,
D.J.Hofman$^6$,
R.S.Hollis$^6$,
R.Ho\l y\'{n}ski$^3$,
B.Holzman$^2$,
A.Iordanova$^6$,
E.Johnson$^8$,
J.L.Kane$^4$,
J.Katzy$^{4,6}$,
N.Khan$^8$,
W.Kucewicz$^6$,
P.Kulinich$^4$,
C.M.Kuo$^5$,
W.T.Lin$^5$,
S.Manly$^8$,
D.McLeod$^6$,
J.Micha\l owski$^3$,
A.C.Mignerey$^7$,
R.Nouicer$^6$,
A.Olszewski$^3$,
R.Pak$^2$,
I.C.Park$^8$,
H.Pernegger$^4$,
C.Reed$^4$,
L.P.Remsberg$^2$,
M.Reuter$^6$,
C.Roland$^4$,
G.Roland$^4$,
L.Rosenberg$^4$,
J.Sagerer$^6$,
P.Sarin$^4$,
P.Sawicki$^3$,
W.Skulski$^8$,
S.G.Steadman$^4$,
P.Steinberg$^2$,
G.S.F.Stephans$^4$,
M.Stodulski$^3$,
A.Sukhanov$^2$,
J.-L.Tang$^5$,
R.Teng$^8$,
A.Trzupek$^3$,
C.Vale$^4$,
G.J.van~Nieuwenhuizen$^4$,
R.Verdier$^4$,
B.Wadsworth$^4$,
F.L.H.Wolfs$^8$,
B.Wosiek$^3$,
K.Wo\'{z}niak$^3$,
A.H.Wuosmaa$^1$,
B.Wys\l ouch$^4$\\
\vspace{3mm}
\footnotesize
$^1$~Argonne National Laboratory, Argonne, IL 60439-4843, USA
$^2$~Brookhaven National Laboratory, Upton, NY 11973-5000, USA
$^3$~Institute of Nuclear Physics, Krak\'{o}w, Poland
$^4$~Massachusetts Institute of Technology, Cambridge, MA 02139-4307, USA
$^5$~National Central University, Chung-Li, Taiwan
$^6$~University of Illinois at Chicago, Chicago, IL 60607-7059, USA
$^7$~University of Maryland, College Park, MD 20742, USA
$^8$~University of Rochester, Rochester, NY 14627, USA
}
}
\begin{document}

% typeset front matter
\maketitle

\begin{abstract}
The PHOBOS experiment at RHIC has measured the multiplicity of
primary charged particles as a function of centrality and 
pseudorapidity in Au+Au collisions at $\snn = $ 19.6, 130 and 200 GeV.  
Two kinds of universal behavior are observed in charged particle production
in heavy ion collisions.
The first is that forward particle production, over a range of
energies, follows a universal limiting curve with a non-trivial
centrality dependence.
The second arises from comparisons with $\pp/\pbarp$ and $\epem$ data.  
$\ratio$ in nuclear collisions at high energy scales
with $\s$ in a similar way as $\nch$ in $\epem$ collisions and
has a very weak centrality dependence.
This feature may be related to a reduction in the leading
particle effect due to the multiple collisions suffered per participant
in heavy ion collisions.

\end{abstract}

\section{Introduction}
The PHOBOS experiment has measured $\dndeta$ and the
average multiplicity of charged particles $\ntot$ produced in heavy ion
collisions for center of mass energies in the nucleon-nucleon
center of mass system, $\snn$, of 19.6, 130 and 200 GeV.
The data is also binned 
as a function of event centrality (impact parameter) characterized
by the number of participating nucleons, $\np$, 
allowing comparisons to elementary systems, like 
$\pp/\pbarp$ and $\epem\rightarrow{\rm hadrons}$.

The PHOBOS multiplicity detector consists of several arrays of silicon
detectors which cover nearly the full solid angle for collision
events.
The ``Octagon'' detector surrounds the interaction region
with a cylindrical geometry below $|z|<50$ cm, covering $|\eta|<3.2$.  
Two sets of 3 ``Ring'' detectors are placed far forward and backward
of the interaction point and surround the beam pipe, covering $3<|\eta|<5.4$.
%The event centrality for 130 and 200 
%GeV is characterized by the multiplicity of charged
%particles measured by two sets of 16 paddle counters covering $3<|\eta|<4.5$.
The methods used for measuring the multiplicity of charged particles 
as well as for extracting $\avenp$ 
has been described in more detail in Ref. \cite{phobos_limfrag}.

\section{Limiting Behavior in Pseudorapidity Distributions}

\begin{figure}[t]
\begin{minipage}{75mm}
\includegraphics[width=8cm]{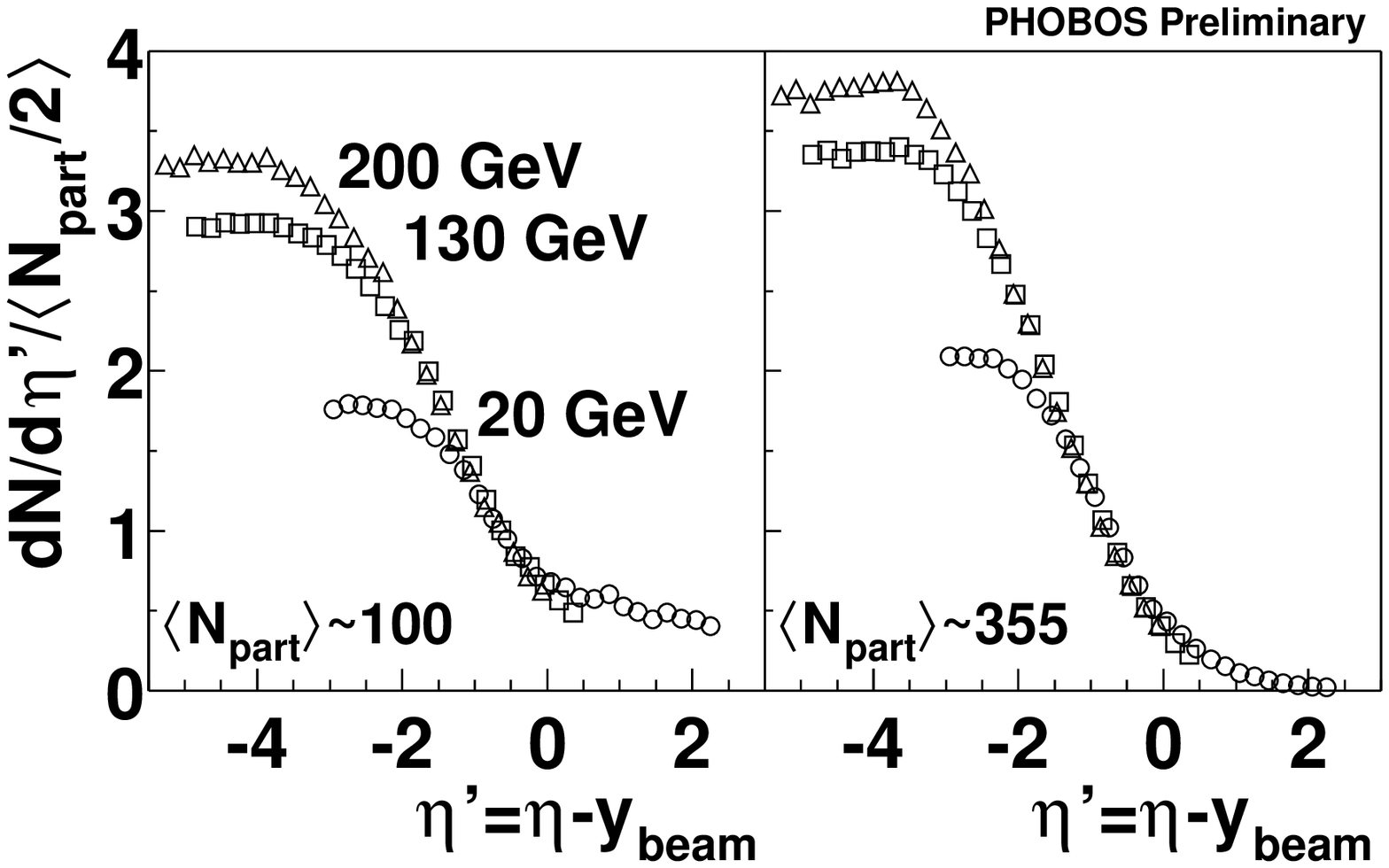}
\caption{$dN/d\etap/\halfnp$ for peripheral 
and central events at three RHIC energies.
\label{fig:limfrag}}
\end{minipage}
\hspace{\fill}
\begin{minipage}{75mm}
\includegraphics[width=80mm]{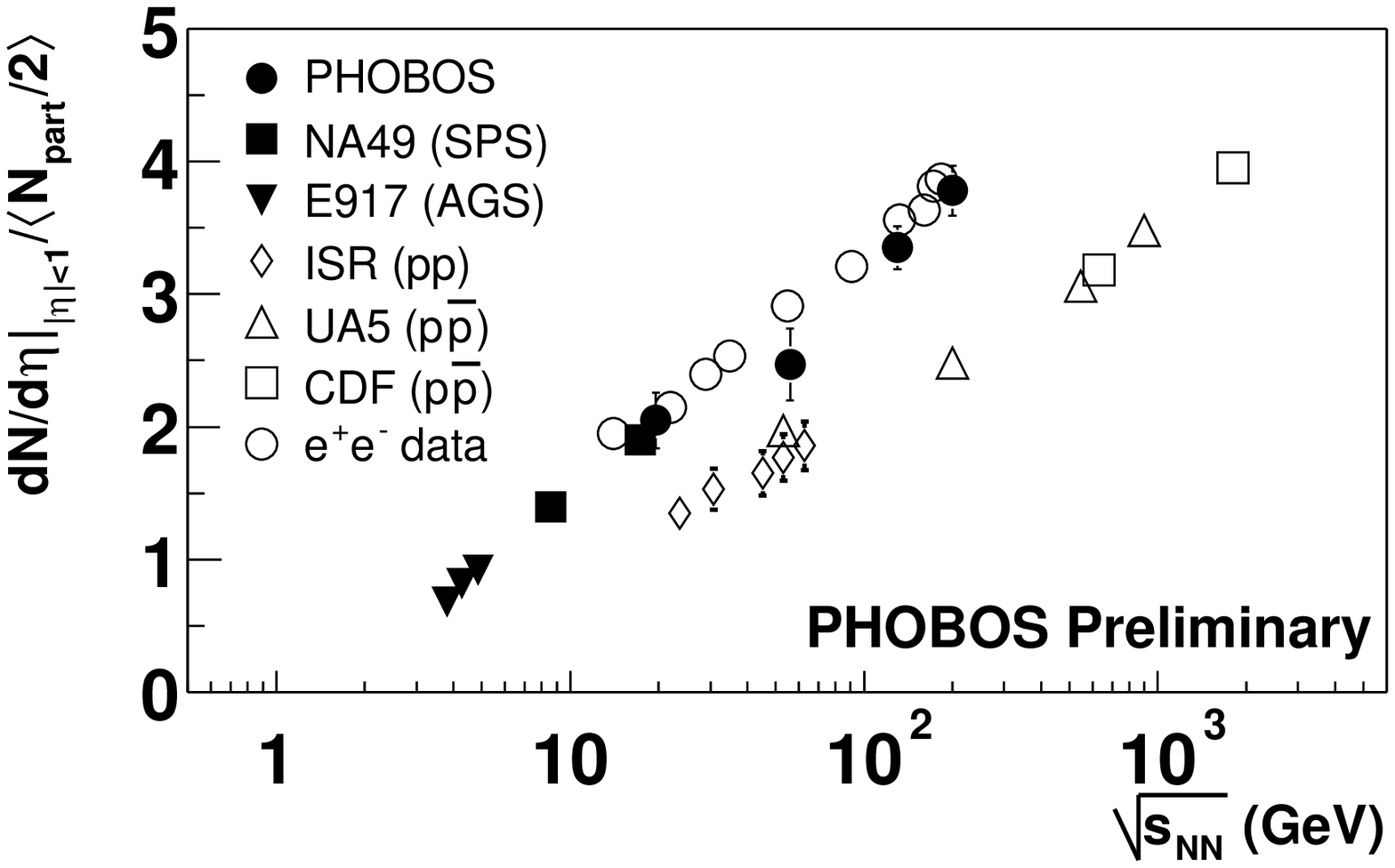}
\caption{
Particle density at midrapidity for A+A, $\pp/\pbarp$
and $\epem$.
\label{fig:midrap}
}
\end{minipage}
\vspace*{-1cm}
\end{figure}

Fig. \ref{fig:limfrag} shows $\dndetapnp$ ($\etap = \eta-\yb$) 
measured at three different
RHIC energies for peripheral ($\avenp\sim100$) and central events ($\avenp\sim355$),
in the left and right panels, respectively.
These show a clear ``limiting
behavior'' in the fragmentation region.
That is, the distributions are independent of beam energy over a substantial
range in $\etap$.
As the beam energy increases, $dN/d\etap$ follows the universal trend
until it reaches 85-90\% of its maximum value at midrapidity, at which
point it stops following the trend.
Similar behavior has been observed in elementary collisions as well,
both in $\pbarp$ collisions \cite{ua5} 
and in $\epem$ annihilation to hadrons \cite{delphi}.

The limiting curve constrains the energy dependence of 
the charged particle multiplicity.
It also 
varies with centrality in such a way that the increases seen at low $\etap$ 
(which is midrapidity in $\eta$) as $\np$ increases, 
are accompanied by decreases
near $\etap\sim 0$ (forward rapidities), as seen in Fig. \ref{fig:limfrag}.
It is not clear why this behavior occurs, e.g.  
from energy conservation or some kind of long-range correlation.

\section{Comparison with Elementary Systems}

\begin{figure}[t]
\begin{minipage}[c]{75mm}
%\begin{center}
\includegraphics[height=70mm]{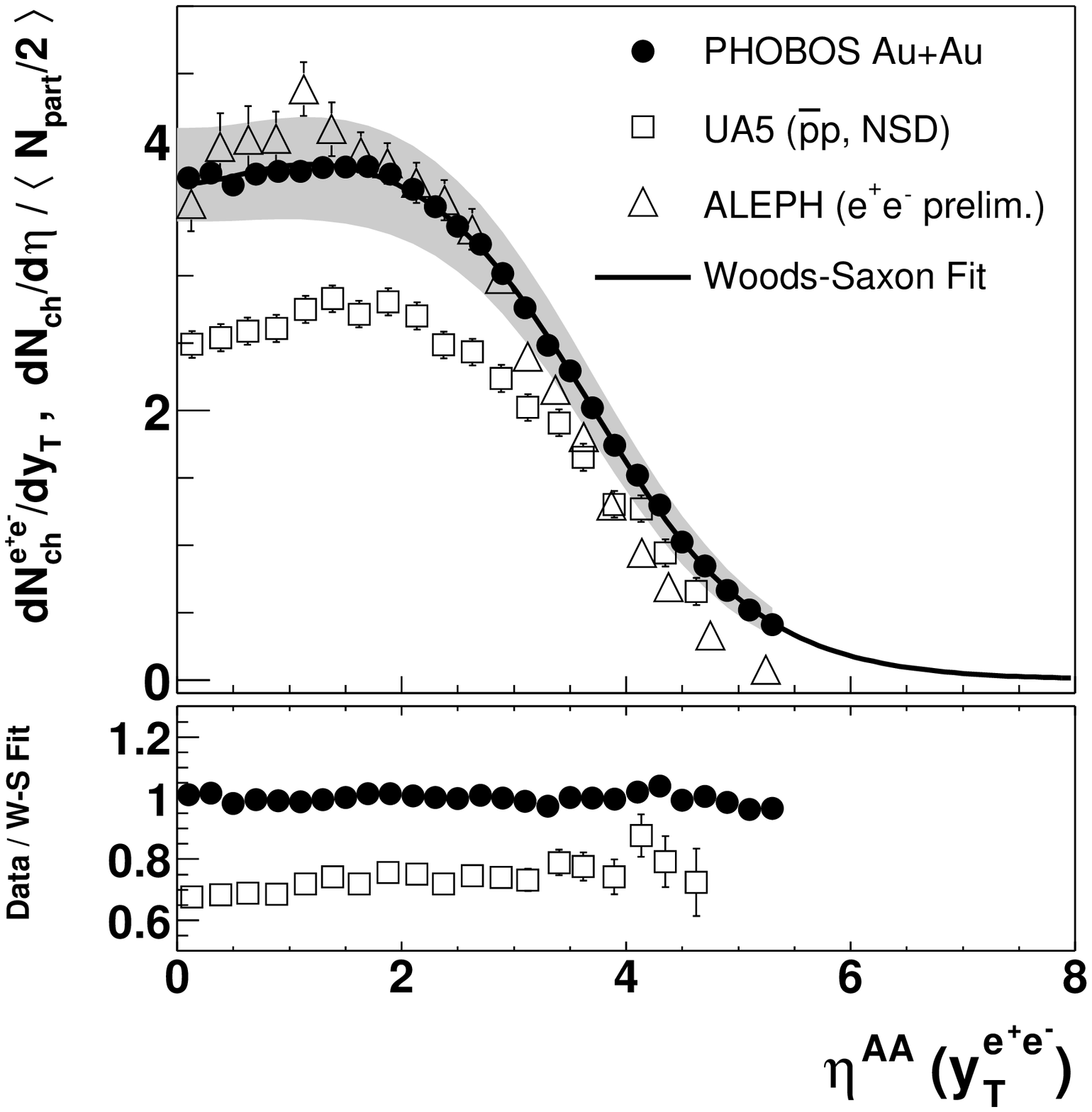}
\caption{
Top: $\dndeta$ for central Au+Au and $\pbarp$
collisions compared with $dN/dy_T$ for $\epem$ data, all at $\snn=$ 200 GeV.
Bottom: Au+Au and $\pbarp$ data divided by a Woods-Saxon fit to the Au+Au data.
\label{fig:AA_ee_pp}
}
%\end{center}
\end{minipage}
\hspace{\fill}
\begin{minipage}[c]{75mm}
%\begin{center}
\includegraphics[height=85mm]{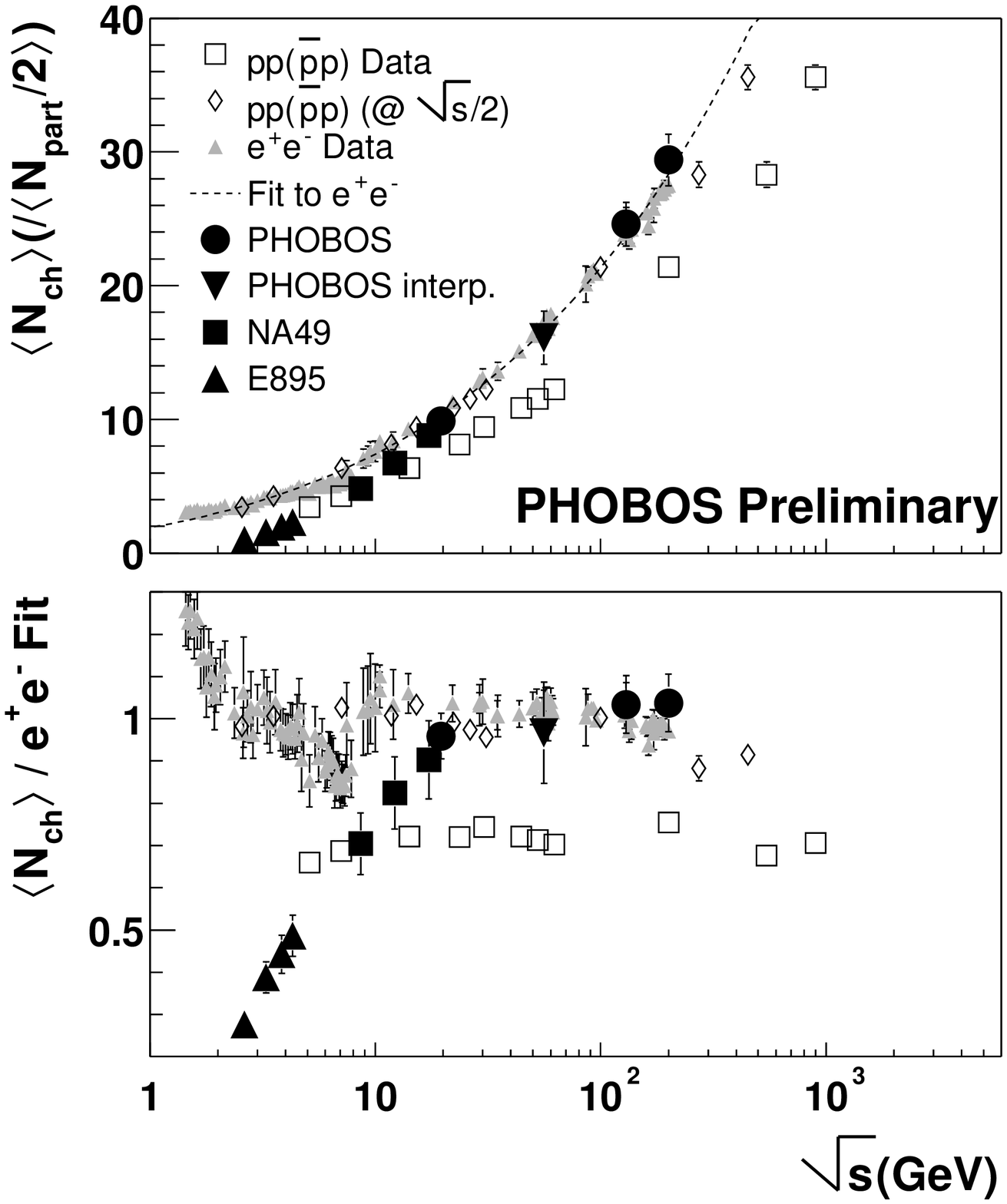}
\caption{
Comparison of $\ratio$ for A+A, $\pp/\pbarp$, and $\epem$
data compared with a fit to the $\epem$ data.
\label{fig:total_ratio}
}
\end{minipage}
\vspace*{-1cm}
%\end{center}
\end{figure}

Comparisons of the plateau height, $\dndetaone/\halfnp$, in heavy ion
collisions with $\pbarp$ data have been made previously \cite{phobos_midrap}.  
However, we can also include data from $\epem$, presented as 
$dN/dy_T$, the rapidity density along the event thrust axis,
calculated assuming the pion mass. 
JETSET calculations suggest that the
difference between $dN/dy_T$ and $dN/d\eta$ is less than $\pm10\%$
for $|y|<4$.  
In the comparison at midrapidity, shown in Fig. \ref{fig:midrap},
we find two interesting features.  First, the energy dependence of
all of the systems is approximately 
logarithmic, at least below 200 GeV.
Secondly, while it has been noticed that heavy ion data is 40-50\%
above $\pp/\pbarp$ data, the $\epem$ data has the same trend and
a similar magnitude (within 10\%) as $AA$ 
over a large range in $\s = 14-183$ GeV \cite{delphi}.
This correspondence holds over the bulk of the angular distribution,
as shown in Fig. \ref{fig:AA_ee_pp}, where central Au+Au (divided by 
$\halfnp$), $\pbarp$ \cite{ua5} and $\epem$ \cite{ALEPH}
data are compared, all at $\s=200$ GeV.  
In the lower panel, we observe
that the shapes of the Au+Au and $\pbarp$ distributions
are also very similar over
a large $\eta$ range, but the integrals differ by $\sim40\%$.

In Fig. \ref{fig:total_ratio}, we compare $\ratio$ in heavy ion 
collisions \cite{heavy-ion}
to $\epem$ and $\pp/\pbarp$ data over a large range in $\s$
\cite{Groom:in}.
It is observed that $\ratio$ lies below $\pp$ at low energies,
passes through the $\pp$ data around $\s\sim 10$ GeV, and then 
gradually joins with the $\epem$ trend above CERN SPS energies.
These comparisons can be seen more clearly by dividing all of the
data by a fit to the $\epem$ data \cite{Mueller:cq}.
The $\pp/\pbarp$ data follows the same trend as $\epem$, but it can be
shown that it matches very well if the ``effective energy'' $\seff = \s/2$
is used, which accounts for the leading particle effect seen
in $\pp$ collisions \cite{basile}.  
Ref. \cite{basile} finds that bulk particle
production in $\pp$ and $\epem$ data
does not depend in detail on the collision system but 
rather the energy available for particle production.
In this scenario, the Au+Au data suggests a substantially reduced leading
particle effect in central collisions of heavy nuclei at high energy.

%The energy dependence of the heavy ion data might be an effect of the
%non-vanishing baryon density at lower energies.
The alleviation of the leading particle effect 
might not be so surprising in nuclear collisions.
Each participating nucleon is struck $\nubar > 3$
times on average as it passes through the oncoming gold nucleus
for $\np>65$.  One could speculate that the multiple
collisions transfer much more of the initial longitudinal energy into
particle production.
%the forward direction towards midrapidity.
This naturally leads to the scaling of total particle production
in heavy ion collisions with $\np$, as seen in Fig. \ref{fig:ntot},
reminiscent of the ``wounded nucleon model'' \cite{wounded}
but with the scaling factor determined by $\epem$ rather than $\pp$.

%The alleviation of the leading particle effect 
%might not be so surprising in central nuclear collisions.
%In the Glauber model, each participating nucleon is typically
%struck $\nubar \sim 6$
%times on average as it passes through the oncoming gold nucleus
%in a central event.  One could speculate that the multiple
%collisions simultaneously 
%excite and dissociate the participating nucleons, 
%transferring nearly all of the energy from
%the forward direction towards midrapidity.
%This naturally leads to the scaling of total particle production
%in heavy ion collisions with $\np$, as seen in Fig. \ref{fig:ntot},
%but with the scaling factor determined by $\epem$ rather than $\pp$.
%This scaling is not substantially modified even 
%in peripheral events ($\np=65$, which corresponds to $\nubar \sim 3$).

%\vspace*{-2mm}

\section{Conclusions}
\begin{floatingfigure}[r]{8cm}
\begin{center}
\includegraphics[width=70mm]{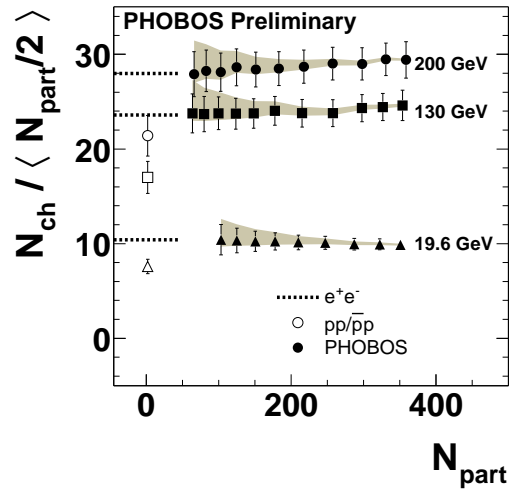}
\caption{
The total number of charged particles per participant pair 
shown as
a function of $\np$ for $\snn = $ 19.6, 130, and 200 GeV.
\label{fig:ntot}
}
\end{center}
\end{floatingfigure}

In conclusion, PHOBOS has observed two kinds of universal behavior. 
The first is an energy-independent, but centrality-dependent, 
universal limiting distribution
of charged particle production away from midrapidity.
%
%The second is that the total particle production is consistent
%with an effective energy per participant pair of approximately
%$\s$ (rather than $\s/2$ seen in $\pbarp$) that fragments into 
%a similar number of particles as $\epem$ annihilations.  
The second is that the total charged particle multiplicity
per participant pair in heavy ion collisions above CERN SPS
energies scales with $\s$ in a similar way as $\epem$ collisions.
These two kinds of universality strongly constrain the energy
dependence of the total charged particle multiplicity and angular
distributions, and may offer a new perspective on particle production
in heavy ion collisions.

\section{Acknowledgements}
{
\footnotesize
This work was partially supported by U.S. DOE grants DE-AC02-98CH10886,
DE-FG02-93ER40802, DE-FC02-94ER40818, DE-FG02-94ER40865, DE-FG02-99ER41099, and
W-31-109-ENG-38 as well as NSF grants 9603486, 9722606 and 0072204.  The Polish
group was partially supported by KBN grant 2 PO3B 10323.  The NCU group was
partially supported by NSC of Taiwan under contract NSC 89-2112-M-008-024.
}

%\vspace*{-2mm}

\end{document}